# Nonlinear Topological Valley Hall Edge States Arising from Type-II Dirac Cones

Hua Zhong[1,#], Shiqi Xia[2,#], Yongdong Li[1], Yiqi Zhang[1,*], Daohong Song[2,*], Chunliang Liu[1], and Zhigang Chen[2,3,*]

[1]*Key Laboratory for Physical Electronics and Devices of the Ministry of Education & Shaanxi Key Lab of Information Photonic Technique, School of Electronic Science and Engineering, Faculty of Electronic and Information Engineering, Xi'an Jiaotong University, Xi'an 710049, China*
[2]*MOE Key Laboratory of Weak-Light Nonlinear Photonics, TEDA Applied Physics Institute and School of Physics, Nankai University, Tianjin 300457, China*
[3]*Department of Physics and Astronomy, San Francisco State University, San Francisco, CA 94132, USA*
[#]*The authors contribute equally to this work*
[*]*Corresponding authors: zhangyiqi@xjtu.edu.cn, songdaohong@nankai.edu.cn, zhigang@sfsu.edu*

**Abstract:** Type-II Dirac/Weyl points, although impermissible in particle physics due to Lorentz covariance, were uncovered in condensed matter physics [1-4], driven by fundamental interest and intriguing applications of topological materials. Recently, there has been a surge of exploration of such generic points using various engineered platforms including photonic crystals [5], waveguide arrays [6,7], metasurfaces [8], magnetized plasma [9] and polariton micropillars [10], aiming towards relativistic quantum emulation and understanding of exotic topological phenomena. Such endeavors, however, have focused mainly on linear topological states in real or synthetic Dirac/Weyl materials. Here, we demonstrate nonlinear valley Hall edge states (VHESs) in laser-writing anisotropic photonic lattices hosting type-II Dirac points [11,12]. These self-trapped VHESs, manifested as topological gap quasi-solitons, are fundamentally distinct from their linear counterparts in type-I lattices and from all previously found topological solitons. Our finding may provide a route for understanding nonlinear phenomena in Lorentz-violating topological systems and for developing advanced light sources from topological insulator lasers [13-16].

**Keywords:** photonic topological insulator, type-II Dirac cone, valley Hall edge soliton, modulation instability, inversion-symmetry breaking, Berry curvature, domain wall

The concept of topological phase was introduced from condensed matter physics into the realm of photonics about a dozen years ago [17,18], which has since led to the burgeoning development in photonic topological insulators (PTIs) [19] and topological photonics in general [20]. Recently, nonlinear topological photonics has attracted increasing attention [21], as nonlinearity exists inherently in many photonic topological systems such as topological insulator lasers [13-16]. Topological edge solitons, for example, have been proposed to exist in several PTI settings [22-26], which typically require breaking of the time-reversal symmetry (TRS) by nontrivial modulation or applying external magnetic fields, posing challenges for their experimental realization. Indeed, the only observation of topological solitons made thus far was in the bulk of an anomalous PTI [27], which relied on a Kerr nonlinearity and periodic driving of anomalously coupled waveguide arrays. Such Floquet solitons in a topological bandgap exhibit only local cyclotron-like motion without dynamical transport.

It is natural to ask if topological edge solitons can exist in TRS-preserved systems, as transport of linear and nonlinear edge states in PTIs without complex modulation or external magnetic fields is essential for many applications. For instance, coupling and locking of semiconductor laser arrays to produce coherent high-power laser sources is one of the key motivations behind the development of topological insulator lasers [13-16], for which the most promising approach could rely on PTI symmetry design [28] and the valley Hall effect [16,29,30] without breaking the TRS. In particular, a nonlinear valley Hall edge (VHE) state in Dirac systems may inherit topological protection from its linear counterpart [31,32], thus benefiting from easy implementation without any need of time modulation or external magnetic fields.

A general consensus is that many topological and valley Hall effects are mediated by materials with nontrivial degeneracy in momentum space, characterized by Dirac/Weyl cones. In the two-dimensional case, a well-celebrated example is graphene, in which electrons around the Dirac cones behave as massless Dirac fermions. Different from type-I Dirac cones in graphene where the corresponding Fermi surface is a point, there are also so-called tilted type-II Dirac cones with the Fermi surface being a pair of crossing lines [10]. Type-II Dirac cones violate the Lorentz symmetry, thus allowing for quasi-particle-mediated phenomena that do not exist in high energy physics analogically investigated in the classical platforms. In photonics, for example, type-II Dirac cones are expected to bring new features due to their non-isotropic transport properties and distinctive dispersions [7,8,11]. Photonics lattices can be readily designed to possess such Dirac cones [12] without sophisticated material engineering or complex modulation on the lattices, making them an attractive platform for light manipulation and investigation of nontrivial topological phenomena arising from type-II Dirac cones.

In this Letter, we propose a scheme to establish type-II Dirac photonic lattices in a nonlinear medium, and thereby unveil the existence of topologically protected nonlinear VHE states. Specifically, by properly stretching a dislocated Lieb lattice, we realize type-II Dirac points in the momentum space. Then, by judiciously breaking lattice inversion symmetry, we build up a domain wall (DW) between two type-II lattices with opposite Berry curvatures around the corresponding Dirac points, where we uncover a family of nonlinear VHE states. Theoretically, we show that it is essential to have the type-II Dirac lattices for formation of such nonlinear localized VHE states, for which self-focusing nonlinearity balances normal dispersion in a large range that cannot be satisfied in conventional type-I lattices. Robust transport of VHE quasi-solitons mediated by modulation instability (MI) is also demonstrated by numerical simulations. Experimentally, we fabricate the type-II lattices by direct cw-laser-writing and we observe both linear and nonlinear VHE states at the DW.

An exemplary type-II Dirac photonic lattice is displayed in Fig. 1(a) with $\delta_l = 4$, and its Brillouin zone (BZ) spectrum is shown in Fig. 1(b), in which the first BZ is indicated by the shaded hexagon. There are three sites in each unit cell of the type-II lattice, and they are labeled as A, B and C in Fig. 1(a). Such a lattice can be considered as the outcome of properly stretching of a dislocated Lieb lattice in the vertical direction [12]. The type-II photonic lattice is also experimentally established by cw-laser-writing (see Methods) in a bulk nonlinear photorefractive crystal, and a typical lattice structure and measured BZ spectrum is shown in Fig. 1(c) and Fig. 1(d), respectively. The experimental results in Figs. 1(c,d) match perfectly the numerical results in Figs. 1(a,b). We solve for the band structure of the type-II lattice based on Eq. (2) – the dimensionless version of Eq. (1) (see **Methods**), and the result is shown in Fig. 1(e). Evidently, under proper lattice design, the bands are tilted and connected at some points - the type-II Dirac points, which are induced solely by the spatial symmetry of the lattice [12]. To see more clearly, we zoom in one Dirac cone indicated by a red circle in Fig. 1(e), and let one horizontal plane (corresponding to the Fermi surface) go across the Dirac point. The intersection at the Dirac cone forms a pair of crossing lines in the momentum space (see the inset in Fig. 1), showing the characteristic feature of the type-II Dirac cone [10].

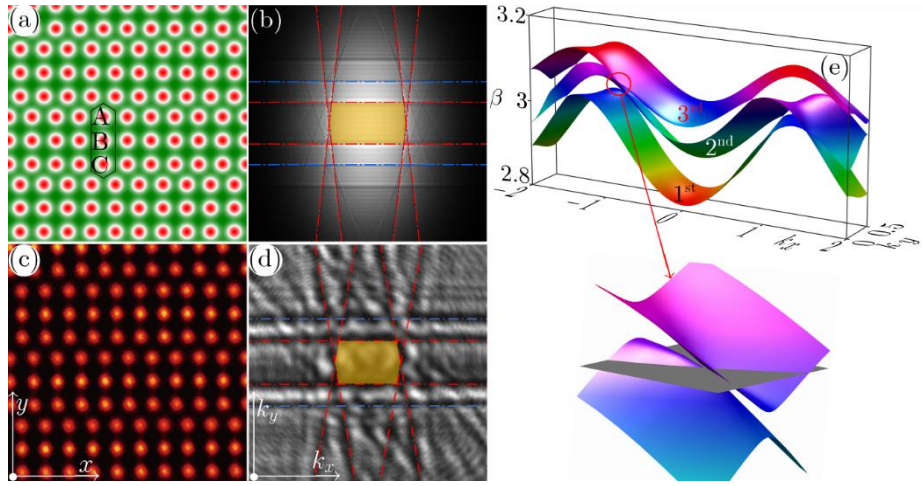

**Fig. 1. Photonic lattices with type-II Dirac cones.** (a) Numerically constructed lattice structure with A, B and C marked as the three sites in a unit cell indicated by a hexagon. (b) Brillouin zone spectrum with the red dashed lines contouring the first BZ (shaded region). (c,d) Experimental realization of the lattice and its BZ spectrum obtained with white-light BZ spectroscopy technique. (e) Band structure of the photonic lattice, in which type-II Dirac points exist due to the tilted band-touching of both 1st and 2nd bands, and 2nd and 3rd bands. The inset is a zoom-in plot of the type-II Dirac cone indicated by a red circle. The gray horizontal plane in $(k_x, k_y)$ coordinate goes across the Dirac point and intersects with the Dirac cone to form a pair of crossing lines – a pristine direct manifestation of the type-II Dirac cone.

In Figs. 1(a) and 1(c), the depths of all lattice sites are identical, but if we intentionally increase the depth of one site in the unit cell, the inversion symmetry of the lattice will be broken. We first increase the depth of site A from $\delta_A = 4$ to $\delta_A = 4.5$, as shown in Fig. 2(a), and we find that the band crossings in the band structure indeed disappear as the gap opens. We also display the calculated Berry curvature $\Omega$ of the first band in the $(k_x, k_y)$ plane with the yellow and blue colors representing the positive and negative values. One can calculate the total Berry curvature which is zero over the first BZ, however, the Berry curvature is nonzero at each valley. Therefore, the valley Chern number $C_v = \frac{1}{2\pi} \iint_{\text{valley}} \Omega(k_x, k_y) \, dk_x dk_y$ can be numerically obtained for each valley, which is $C_v = \pm 1/2$. In Fig. 2(c), we break the inversion symmetry of the lattice by increasing the depth of site C from $\delta_C = 4$ to $\delta_C = 4.5$. The corresponding band structure in Fig. 2(d) is same as that in Fig. 2(b), but the Berry curvature of the first band is opposite. Therefore, a DW can be established between two lattices with different breaking-inversion-symmetries, as shown in Fig. 2(e), where the DW is outlined by a rectangle. Across the DW, the difference of the valley Chern numbers is $|\Delta C_v| = |\pm 1/2 - (\mp 1/2)| = 1$, which indicates that a topological nontrivial edge state emerges along the DW according to the bulk-edge correspondence principle. Note that the lattice in Fig. 2(e) is periodic along $x$ direction (12 periods are shown), but with boundaries along $y$ axis. The corresponding band structure $\beta(k_x)$ can be numerically obtained by using the plane-wave expansion method, which is displayed in Fig. 2(f). The black curves in the band structure represent the bulk states, while the red curve in the band gap is the VHE state that distributes along the DW. Similar to previously reported VHE states in type-I Dirac photonic lattices [29,30], the VHE state here is also non-monotonous, which means that the DW supports counter-propagating VHE states in this TRS-preserving lattice. In Fig. 2(g), the amplitude profiles of the VHE state at four selective Bloch momenta are shown. Clearly, the edge states are well localized along the DW.

Under the action of self-focusing nonlinearity, it is expected that the VHE soliton, if exists, should reside in the normal dispersive regime. We thus solve for the first-order derivative $\beta'$ and second-order derivative $\beta''$ of the linear VHE state in Fig. 2(f), and the results are displayed in Fig. 3(a). Indeed, the VHE state may move along positive or negative $x$ direction, because its group velocity determined by the sign of $\beta'$ can be positive or negative, in accordance with the TRS-preserving characteristics. (Here the VHE state with $\beta' > 0$ moves along positive $x$ direction). Notably, as shown in Fig. 3(a), the normal dispersion ($\beta'' > 0$) range is quite large in the first BZ: $-0.235 \leq k_x/K \leq 0.235$, where K defines the width of the first BZ. Therefore, it is reasonable to search for nonlinear VHE states and investigate their transport properties at the Bloch momentum around $k_x/K = 0.1$. Indeed, we found a family of such nonlinear solutions: the peak amplitude $a = \max\{|\psi|\}$ and power $P = \iint |\psi|^2 dxdy$ of the nonlinear VHE states at $k_x/K = 0.1$ are plotted in Fig. 3(b) as a function of the nonlinearity-dependent propagation constant $\mu$ (or "energy shift") (see **Methods**). One finds that both $a(\mu)$ and $P(\mu)$ decrease as $\mu$ increases, reducing to nearly zero at $\mu \sim 2.741$ where the nonlinear localized states resemble or reduce to the linear ones. This is quite direct, because a nonlinear edge state typically bifurcates from corresponding linear counterpart under the action of nonlinearity, with inherited topological feature. We note that the nonlinear VHE states found at the DW for a large range of power shown in Fig. 3(b) are nearly identical in intensity and phase structures. To avoid mixing with the bulk states, we consider only the nonlinear VHE states residing in the band gap between the linear VHE state [red curve in Fig. 2(f)] and bulk states [bottom black curves in Fig. 2(f)], e.g., with corresponding propagation constant $\beta \sim 2.65$.

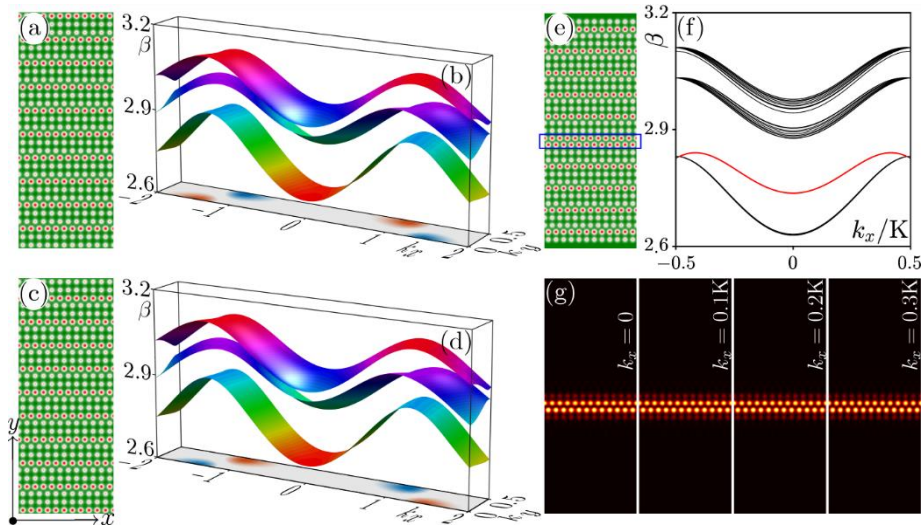

**Fig. 2. Linear topological VHE states at the domain wall between two Type-II Dirac photonic lattices.** (a) Inversion-symmetry-broken photonic lattice with depth of site A increased. (b) Band structure corresponding to the lattice in (a) with the Berry curvature of the first band displayed in the $(k_x, k_y)$ plane. (c,d) Same layout as in (a,b), but with depth of site C increased instead. The values of the Berry curvatures in (b) and (d) are opposite. (e) A DW is established in the photonic lattice formed by combining (a) and (c), highlighted by a blue rectangle. The lattice is periodic along $x$, but has boundaries

along $y$-direction. (f) Band structure of the lattice in (e); the red curve represents the topologically protected VHE states along the DW. (g) Profiles of VHE states at selective Bloch momenta.

Now, we study the robustness of the nonlinear VHE states by examining the development of MI under the action of self-focusing nonlinearity, which is possible since $\beta'' > 0$ at $k_x/K = 0.1$. To do so, we add a random noise to the nonlinear VHE state obtained at $\mu = 2.736$, with a noise amplitude about 5% of the nonlinear edge state shown in Fig. 3(c). Representative propagation of the peak amplitude $a = \max\{|\psi|\}$ of the perturbed state is shown in Fig. 3(d). The amplitude profiles of the nonlinear edge state at two selected distances [marked by two red dots in Fig. 3(d)] are displayed in Fig. 3(c), showing growth of MI during propagation which leads to formation of quasi-soliton filaments along the DW. These MI-induced soliton filaments are considered as precursors for the formation of topological VHE gap solitons. Without loss of generality, we take out one of such bright filaments at $z = 1750$ [indicated by a green circle in Fig. 3(c)] as the input and investigate its long-distance propagation dynamics. To this end, we introduce another physical quantity – the barycenter of the filament defined as $x_c = P^{-1} \iint x|\psi|^2 dx dy$ – to record its movement during propagation. We first shift the selected filament in Fig. 3(c) to the center of the window ($x = 0$) and then track its propagation. The peak amplitude and barycenter of the filament are displayed in Fig. 3(e), showing the stability of a moving VHE quasi-soliton. Even over an extremely long propagation ($z \sim 32000$), the peak amplitude remains nearly invariant and the barycenter exhibits a saw-teeth-like oscillating behavior. (The appearance of the oscillation is mainly due to the simulation window: for a chosen window along $x$, the movement of the VHE quasi-soliton appears at the left end of the window when it reaches the right end, resulting in apparent periodic jump of the barycenter between two ends). In Fig. 3(f), snapshots of the quasi-soliton taken at different propagation distances are displayed. We observes clearly that the quasi-soliton moves along positive $x$-direction with a constant speed, and it remains localized with negligible radiation loss either along the DW or into the bulk – a result of interplay between nonlinearity and topological protection! For direct comparison, we propagate the same input filament in the linear lattice, i.e., removing the nonlinear term $|\psi(x,y,z)|^2$ in Eq. (3). As expected, without the balance from the nonlinearity, the filament spreads quickly along the DW because of diffraction [see Fig. 3(g)], yet remains localized in the direction perpendicular to the DW.

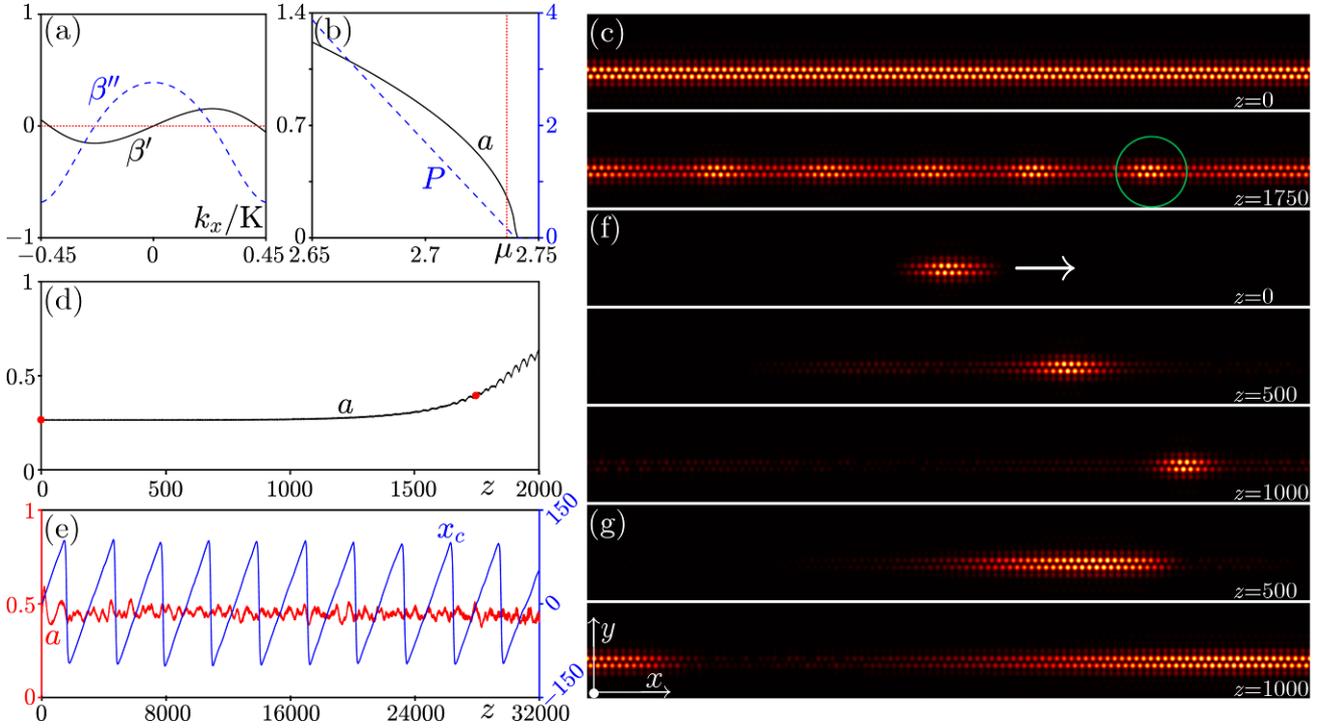

**Fig. 3. Nonlinear topological VHE states and robust transport of quasi-solitons from modulation instability.** (a) Dispersion spectrum of the linear edge states in Fig. 2(f). Solid curve is for $\beta' = d\beta/dk_x$, and dashed curve for $\beta'' = d^2\beta/dk_x^2$. The red dotted line shows the zero value of the derivatives. VHE quasi-solitons are found only in the region with $\beta'' > 0$. (b) Plots of amplitude $a$ and power $P$ of the nonlinear edge states, reducing to linear states at $k_x = 0.1K$, versus nonlinear energy shift $\mu$. The red dotted line corresponds to $\mu = 2.736$. (c) Profiles of a typical nonlinear edge state (superimposed with 5% random noise) found at $\mu = 2.736$ at $z = 0$ and $z = 1750$. (d) Amplitude of the nonlinear state in (c) versus propagation distance. (e) Amplitude $a$ (red curve) and barycenter $x_c$ (blue curve) of a quasi-soliton filament (marked by the green circle in (c)) resulting from MI, showing robustness of the quasi-soliton over extremely long distances, as seen also in the intensity pattern snapshots (f) during transport. (g) Spreading of the same input as in (f) at $z = 0$ along the DW during linear propagation for comparison. Notice that the energy does not radiate into the bulk due to valley Hall topological protection. (Here $z = 1000$ corresponds to a physical distance 4 meters).

In the experiment, with the cw-laser-writing technique [33] used already for making the type-II Dirac photonic lattice shown in Fig. 1(c), we establish two such lattices but with different broken inversion-symmetry and a DW between them. Such a structure [corresponding to Fig. 2(e)] is shown in Fig. 4(a), where the DW is marked by a white solid line. To examine the nonlinear VHE states formation presented in Fig. 3(e), two out-of-phase broad elliptical Gaussian beams are employed as a probe, with its launching position marked by the white dashed oval in Fig. 4(a). The linear output intensity patterns in real space and their corresponding spectra in momentum space for three different Bloch momenta $k_x/K = -0.3, 0, +0.3$ are shown in Figs. 4(b-d). We observe that the probe beam is localized in vertical direction (not spreading into the bulk) but somewhat extended along the DW after 1 cm of propagation due to the excitation of the linear VHE state. Interestingly, we find that the output patterns moved slightly to the left (right) comparing to the initial probe position in Fig. 4(b1) [Fig. 4(d1)] for the Bloch momentum $k_x/K = -0.3 (+0.3)$, thanks to the initial transverse velocity, while the output position for $k_x = 0$ remains invariant [Fig. 4(c1)]. Bragg-reflected components are clearly visible in the momentum space spectra [Figs. 4(b2,d2)], in accordance with the BZ of our fabricated type-II lattice [see Fig. 1(d)]. The corresponding nonlinear outputs are shown in Figs. 4(b3-d3) and Figs. 4(d3-d4), which are different from those linear counterparts. Due to the action of self-focusing nonlinearity, the output patterns are now more localized along the DW direction in the real space, and at the same time the spectra more expanded in the momentum space comparing with the linear outputs. These experimental results are corroborated by corresponding simulations shown in Fig. 4(e) with the same parameters. Our results indicate that the self-trapped nonlinear VHE states indeed exist in this type-II Dirac photonic lattice. Due to the limited propagation distance (as set by the crystal length), it is not feasible to experimentally show the long-distance transport of the quasi-solitons as demonstrated in our theoretical analysis.

We have thus demonstrated nonlinear VHE states and formation of quasi-solitons through theoretical analysis and experimental realization of photonic lattices exhibiting type-II Dirac points. We have shown that it is crucial to have the type-II Dirac dispersion and a DW from lattices of opposite Berry curvatures in order to achieve self-trapping of VHE solitons in TRS-preserving topological systems. Our results may provide insights to investigating nonlinear topological phenomena in other type-II Dirac/Weyl systems beyond photonics, which have so far gone AWOL in experiment. From the viewpoint of applications, the concept developed here with nonlinear VHE states could be useful for topological insulator lasers [13-16], where nonlinear topologically protected transport of edge modes could be used to lock many semiconductor emitters to generate coherent high-power laser sources.

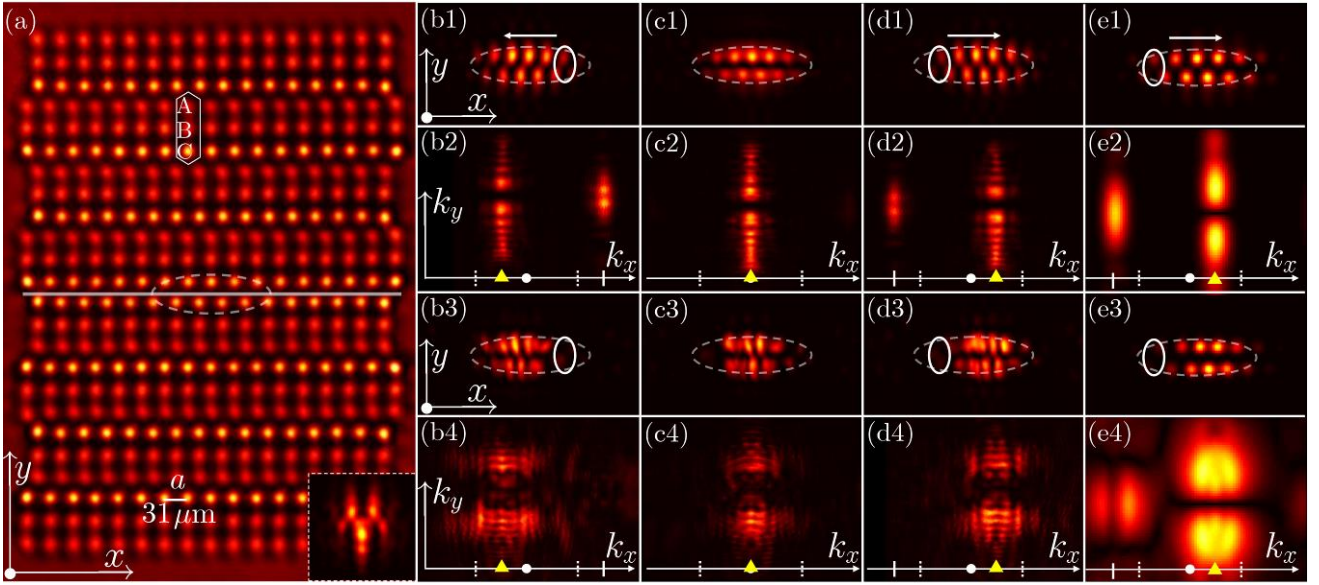

**Fig. 4. Experimental observation of nonlinear topological VHE states.** (a) The type-II lattice with a center DW (marked by the white line) corresponding to Fig. 2(e) established by site-to-site cw-laser-writing technique. The inset shows discrete diffraction from single-site excitation. (b-d) Linear (top two rows) and nonlinear (bottom two rows) outputs in real and momentum space obtained from two out-of-phase broad elliptical beams excitation at the DW (circled by dashed line in (a)). The excitation directions in $k$-space in (b), (c) and (d) are $-3\pi/5a$, $0$, $3\pi/5a$, respectively, marked by yellow triangles, and the white dots mark the BZ center for reference. Dashed and solid ellipses in (b1) mark the initial input position and the portion of light that is diminished at the nonlinear output in (b3), respectively, indicating nonlinearity-induced self-trapping and transport of the VHE state when the initial excitation is tilted to the left. (c,d) The layout is the same as (b) but with different excitation directions. (e1-e4) Simulation results corresponding to (d1-d4). Bragg reflection （row 2 and nonlinear spectral reshaping (row 4) are noticeable from the momentum space.

## Methods

**Experimental setup.** The type-II Dirac photonic lattice is induced in a photorefractive nonlinear crystal

(SBN:61) by adopting the site-to-site cw-laser-writing technique [33] with laser operating wavelength at $\lambda_0 = 532$ nm. The dimension of the crystal is $5 \times 5 \times 10$ mm$^3$. The refractive index change of the sublattices, i.e. depth of site $\delta_{A,B,C}$, is well controlled by the writing time, and it is typically on the order of $10^{-4}$. To excite the VHE state, two out-of-phase broad elliptical Gaussian beams are prepared via a spatial light modulator to cover about 8 lattice sites of the DW, as marked by the white dashed ellipse in Fig. 4(a). The self-focusing nonlinearity is achieved by using a bias electric field of about 70 kV/m, with input power of the probe beam about only 2 $\mu$W.

**Theoretical Model.** The propagation of a light beam in a photonic lattice written in a nonlinear crystal is described by the following Schrödinger-like paraxial wave equation:

$$i\frac{\partial \psi(x,y,z)}{\partial z} = -\frac{1}{2k_0}\left(\frac{\partial^2}{\partial x^2} + \frac{\partial^2}{\partial y^2}\right)\psi(x,y,z) - \frac{k_0 \Delta n(x,y)}{n_0}\psi(x,y,z), \quad (1)$$

where $k_0 = 2n_0\pi/\lambda_0$ is the wave number in the crystal with refractive index $n_0$. For the specific photorefractive SBN:61 crystal used in our experiment, we take

$$\Delta n(x,y) = -\frac{1}{2}n_0^3 \gamma_{33} E_0 \frac{1}{1+I(x,y)}$$

with $I(x,y)$ being the intensity pattern of the lattice beam. The parameters corresponding to experiment are $\lambda_0 = 532$ nm, $n_0 = 2.35$, the bias field $E_0 = 1$ kV/cm, and the electro-optic coefficient $\gamma_{33} = 280$ pm/V. Even with a weak writing beam intensity, the refractive index change reaches typically $\Delta n \sim 1.82 \times 10^{-4}$. While in experiment the lattices are written point by point, the corresponding lattice intensity pattern can be considered as appropriate superposition of Gaussian beams $I(x,y) = \left|\sum_{l=\{A,B,C\},m} \delta_l \exp\{-[(x-x_m)^2 + (y-y_m)^2]/w^2\}\right|^2$, in which $\delta_l$ represents the beam amplitude, $(x_m, y_m)$ is the beam center, and $m$ is an integer. The distance between two nearest-neighbor sites is set to be $d = 30$ $\mu$m and $w = 0.32d$. The diffraction length is $L_z = k_0 r_0^2$ with $r_0 = 12$ $\mu$m representing the typical width of the real incident beam. If we replace $x, y, z$ by $x/r_0, y/r_0, z/L_z$, one obtains the dimensionless version of Eq. (1), which is

$$i\frac{\partial \psi(x,y,z)}{\partial z} = -\frac{1}{2}\left(\frac{\partial^2}{\partial x^2} + \frac{\partial^2}{\partial y^2}\right)\psi(x,y,z) - \frac{k_0^2 r_0^2 \Delta n(x,y)}{n_0}\psi(x,y,z), \qquad (2)$$

The solution of Eq. (2) can be written as $\psi(x,y,z) = u(x,y)\exp(-i\beta z)$ with $\beta$ being the propagation constant (or "linear energy shift") and $u(x,y)$ the Bloch function. Plugging this ansatz into Eq. (2), one can solve for the band structure of the type-II Dirac photonic lattice by applying the plane-wave expansion method in the momentum space $(k_x, k_y)$.

When nonlinearity is considered, the refractive index change will be written as

$$\Delta n(x,y,z) = -\frac{1}{2}n_0^3 \gamma_{33} E_0 \frac{1}{1 + I(x,y) + |\psi(x,y,z)|^2},$$

which is dependent on the beam intensity $|\psi(x,y,z)|^2$ upon propagation. With the nonlinear refractive index change, the solution of Eq. (2) can be still written as $\psi(x,y,z) = u(x,y)\exp(ik_x x - i\mu z)$ if the lattice is strained in $y$ axis with boundaries, with $\mu$ being the "nonlinear energy shift" of the nonlinear topological VHE state. Inserting this ansatz into Eq. (2), the nonlinear topological VHE state can be obtained by using the Newton method.

## Data availability

The data that support the plots within this paper and other findings of this study are available from the corresponding author.

## Code availability

The codes that support the findings of this study are available from the corresponding author upon reasonable request.

## Acknowledgements

This work was supported by the National Key R&D Program of China (2017YFA0303800), the Natural Science Foundation of China (12074308, 11922408, 11674180, U1537210), the Fundamental Research Funds for the Central Universities (xzy012019038, xzy022019076). Y.Z. acknowledges the computational resources provided by the HPC platform of Xi'an Jiaotong University.

## Author contributions

All authors contributed significantly to the work.

## Competing interests

The authors declare no competing interests.

## Additional information

**Supplementary information** is available for this paper at https://doi.org/

**Correspondence and requests for materials** should be addressed to Y.Z., D.S. and Z.C.

**Reprints and permissions information** is available at www.nature.com/reprints.